\documentclass[a4paper]{jpconf} 
\usepackage{graphicx}

\def \rightdownarrow
 {\kern.3em
 \rule[.5ex]{.15mm}{2ex}
 {\mbox{$\kern-0.1em{\longrightarrow}$}}      
 }

\def\lessim{\mathrel {\vcenter {\baselineskip 0pt \kern 0pt  
\hbox{$<$} \kern 0pt \hbox{$\sim$} }}}

\def\gessim{\mathrel {\vcenter {\baselineskip 0pt \kern 0pt   
\hbox{$>$} \kern 0pt \hbox{$\sim$} }}}

\newcommand{\babelle}{\mbox{$B$-Factories}}

\newcommand{\Lumi}{\ensuremath{\mathcal{L}}}			

\newcommand{\lumifb}{\mbox{fb$^{-1}$}}				

\newcommand{\BR}{\ensuremath{\mathcal{B}}}

\newcommand{\tev}{\ensuremath{\mathrm{Te\kern -0.1em V}}}
\newcommand{\gev}{\ensuremath{\mathrm{Ge\kern -0.1em V}}}	
\newcommand{\mev}{\ensuremath{\mathrm{Me\kern -0.1em V}}}	
\newcommand{\kev}{\ensuremath{\mathrm{ke\kern -0.1em V}}}	

\newcommand{\stat}{\ensuremath{\mathit{~(stat.)}}}		
\newcommand{\syst}{\ensuremath{\mathit{~(syst.)}}}		

\newcommand{\CP}{{\rm CP}}                                            

\newcommand{\abd}{\ensuremath{\overline{B}^{0}}}		
\newcommand{\abs}{\ensuremath{\overline{B}^{0}_s}}		

\newcommand{\bn}{\ensuremath{B^{0}_{(s)}}}			

\newcommand{\bdkpi}{\ensuremath{\Bd \to K^+ \pi^-}}

\newcommand{\bskpi}{\ensuremath{\Bs \to K^- \pi^+}}

\newcommand{\cita}[1]{\cite{#1}}

\newcommand{\dedx}{\ensuremath{\mathit{dE/dx}}}

\newcommand{\acp}{\ensuremath{{\cal A}_{\CP}}}
\newcommand{\acpbdkpi}{\ensuremath{\acp(\bdkpi)}}

\def\babar{\mbox{\slshape B\kern-0.1em{\smaller A}\kern-0.1em B\kern-0.1em{\smaller A\kern-0.2em R}}}

\newcommand{\Bhh}{\ensuremath{\bn \rightarrow h^{+}h^{'-}}}

\newcommand{\Bdpipi}{\ensuremath{\Bd \rightarrow \pi^+ \pi^-}}

\newcommand{\BdKpi}{\ensuremath{\Bd \rightarrow K^{+} \pi^-}}

\newcommand{\aBdKpi}{\ensuremath{\abd \rightarrow K^- \pi^+}}

\newcommand{\BsKpi}{\ensuremath{\Bs \rightarrow K^- \pi^+}}

\newcommand{\aBsKpi}{\ensuremath{\abs\rightarrow K^+ \pi^-}}

\newcommand{\BsKK}{\ensuremath{\Bs \rightarrow  K^+ K^-}}

\newcommand{\Bspipi}{\ensuremath{\Bs \rightarrow  \pi^+ \pi^-}}

\newcommand{\BdKK}{\ensuremath{\Bd \rightarrow  K^+ K^-}}

\newcommand{\Lbppi}{\ensuremath{\Lambda_{b}^{0} \rightarrow p\pi^{-}}}

\newcommand{\LbpK}{\ensuremath{\Lambda_{b}^{0} \rightarrow pK^{-}}}

\newcommand{\BsDspi}{\ensuremath{\Bs \rightarrow D^{-}_{s} \pi^+}}

\newcommand{\BsDsK}{\ensuremath{\Bs \rightarrow D^{\mp}_{s} K^{\pm}}}

\newcommand{\ACPddef}{\ensuremath{{\frac{\BR (\aBdKpi)-\BR 
(\BdKpi)}{\BR (\aBdKpi)+\BR (\BdKpi)}}}}
\newcommand{\ACPsdef}{\ensuremath{{\frac{\BR (\aBsKpi)-\BR 
(\BsKpi)}{\BR (\aBsKpi)+\BR (\BsKpi)}}}}
\newcommand{\rateratiodef}{\ensuremath{\frac{\mathit{f_d}}{\mathit{f_s}}\times{\frac{\Gamma (\aBdKpi)-\Gamma 
(\BdKpi)}{\Gamma (\aBsKpi)-\Gamma (\BsKpi)}}}}
\newcommand{\BdpipisuBdKpidef}{\ensuremath{\frac{\BR(\Bdpipi)}{\BR(\BdKpi)}}}
\newcommand{\BsKKsuBdKpidef}{\ensuremath{\frac{\mathit{f_s}}{\mathit{f_d}}\times \frac{\BR(\BsKK)}{\BR(\BdKpi)}}}

\newcommand{\BsKpisuBdKpidef}{\ensuremath{\frac{\mathit{f_s}}{\mathit{f_d}}\times\frac{\BR(\BsKpi)}{\BR(\BdKpi)}}}
\newcommand{\BspipisuBdKpidef}{\ensuremath{\frac{\mathit{f_s}}{\mathit{f_d}}\times\frac{\BR(\Bspipi)}{\BR(\BdKpi)}}}
\newcommand{\BdKKsuBdKpidef}{\ensuremath{\frac{\BR(\BdKK)}{\BR(\BdKpi)}}}

\newcommand{\LbppisuLbpKdef}{\ensuremath{\frac{\BR(\Lbppi)}{\BR(\LbpK)}}}

\def\beq{\begin{equation}}
\def\eeq{\end{equation}}
\def\bea{\begin{eqnarray}}
\def\eea{\end{eqnarray}}

\def\sss{\scriptscriptstyle}

\def\barp{{\raise.35ex\hbox
{${\sss (}$}}---{\raise.35ex\hbox{${\sss )}$}}}
\def\bdbarp{\hbox{$B_d$\kern-1.4em\raise1.4ex\hbox{\barp}}}
\def\bsbarp{\hbox{$B_s$\kern-1.4em\raise1.4ex\hbox{\barp}}}

\def\roughly#1{\mathrel{\raise.3ex\hbox
{$#1$\kern-.75em\lower1ex\hbox{$\sim$}}}}

\newcommand{\acpbskpi}{\ensuremath{\acp(\bskpi)}}

\newcommand{\acpbukpi}{\ensuremath{A_{\rm{\CP}}(B^+ \to K^+ \pi^0)}}

\newcommand{\Bd}{\ensuremath{B^{0}}}

\newcommand{\Bs}{\ensuremath{B_{s}^{0}}}
\newcommand{\Lb}{\ensuremath{\Lambda_{b}^{0}}}

\newcommand{\bear}{\begin{array}}
\newcommand{\ear}{\end{array}}
\newcommand{\bet}{\begin{tabular}}
\newcommand{\eet}{\end{tabular}}
\newcommand{\beqn}{\begin{eqnarray}}
\newcommand{\eeqn}{\end{eqnarray}}

\begin{document}
\title{CP and charge asymmetries at CDF}

\author{Michael Morello for the CDF Collaboration}

\address{University and  I.N.F.N of Pisa - Ed. C,
         Polo Fibonacci, Largo B. Pontecorvo, 3 - 56127 Pisa, Italy}

\ead{michael.morello@pi.infn.it}

\begin{abstract}
We present CDF results on the branching fractions and time-integrated direct \CP\ 
asymmetries for \Bd\ and \Bs\ decay modes into pairs of charmless charged hadrons (pions or kaons). 
We report also the first observation of \BsDsK\ mode and the measurement of its branching fraction. 
\end{abstract}

\section{Introduction}
The interpretation of the CP violation
mechanism is one of the most controversial aspects of the Standard Model.
Many extensions of Standard Model predict that there are new sources of CP violation,
beyond the single Kobayashi-Maskawa phase in the quark-mixing matrix (CKM). Considerations 
related to the observed baryon asymmetry of the Universe imply that such new sources should exist.
The non-leptonic decays of  $B$ mesons 
are effective probes of the CKM matrix and sensitive to these potential new physics effects.
The large production cross section of $B$ hadrons of all kinds at the Tevatron
allows extending such measurements to \Bs\ decays,
which are important to supplement our understanding of \Bd\ meson decays.

The \BsKpi\ mode could be used to measure
$\gamma$~\cite{Gronau:2000md} and its
\CP\ asymmetry could be a powerful model-independent test 
of the source of  direct
\CP\ asymmetry in the $B$ system \cite{Lipkin-BsKpi}. This may provide 
useful information to solve the
current discrepancy between the asymmetries observed in the neutral
\acpbdkpi\ and charged mode \acpbukpi~\cite{HFAG06}. 
A time-dependent, flavor-tagged measurement of \BsDsK\ can provide a measurement of $\gamma$
in a theoretically clean way~\cite{bsdsk_dunietz}.

Throughout this paper, C-conjugate modes
are implied and branching fractions indicate
\CP-averages unless otherwise stated.


\section{Measurements of \Bhh\  decays}

The Collider Detector at Fermilab (CDF) experiment analysed
an integrated luminosity  $\int\Lumi dt\simeq 1$~\lumifb\ sample of pairs of oppositely-charged particles
and reconstructed a sample of 14,500 \Bhh\ decay modes (where $h= K~{\rm or}~ \pi$) after the off-line confirmation
of trigger requirements. 
In the off-line analysis, we chose the selection cuts minimizing the expected uncertainty of the physics
observables to be measured (through 
several ``pseudo-experiments'').
We used two different sets of cuts, respectively optimized
to measure the \CP\ asymmetry \acpbdkpi\ (loose cuts) and to improve the 
sensitivity for discovery and limit setting~\cite{gp0308063} of the not yet observed \BsKpi\ mode.
The resolution in invariant mass and in particle identification (\dedx) is not 
sufficient for separating the individual \Bhh\ decay modes on an event-by-event basis,
therefore we performed a Maximum Likelihood fit which combines kinematic and particle identification information
to statistically determine both the contribution of each mode,
and the relative contributions to the \CP\ asymmetries.
We performed two separate fits: one on the sample selected with loose cuts and one
on the sample selected with tight cuts. 
Significant signals are seen for \Bdpipi, \BdKpi, and \BsKK, previously observed by CDF~\cite{paper_bhh}.
Three new rare modes were observed for the first time \BsKpi, \Lbppi\ and \LbpK,
with a significance respectively of $8.2 \sigma$, $6.0 \sigma$ and $11.5 \sigma$, 
estimated using a $p$-value distribution on pseudo-experiments. 
No evidence was obtained for \Bspipi or  \BdKK\ mode.
\begin{center}
\begin{table}[h]
\centering
\caption{\label{tab:summary}Results on data sample selected with loose cuts (top) and
 tight cuts (bottom). Absolute branching fractions are normalized to the the world--average values
${\mathcal B}(\mbox{\BdKpi}) = (19.7\pm 0.6) \times 10^{-6}$ and
$f_{s}= (10.4 \pm 1.4)\%$ and $ f_{d}= (39.8 \pm 1.0)\%$~\cite{HFAG06}.
The first quoted uncertainty is statistical, the second is systematic.
$N_s$ is the number of fitted events for each mode. For rare modes both systematic and statistical uncertainty on $N_s$ was quoted 
while for abundant modes only the statistical one. For the \Lb\ modes only the ratio \LbppisuLbpKdef\ was measured.}
{\scriptsize
\begin{tabular}{lc|lc|c}
\br
Mode & N$_{s}$ & Quantity & Measurement & \BR (10$^{-6}$)  \\
\mr
\BdKpi         & 4045 $\pm$ 84          &  \ACPddef\         & -0.086 $\pm$ 0.023 $\pm$ 0.009   &                            \\
\Bdpipi        & 1121 $\pm$ 63          & \BdpipisuBdKpidef\ & 0.259 $\pm$ 0.017 $\pm$ 0.016    & 5.10 $\pm$ 0.33 $\pm$ 0.36 \\
\BsKK          & 1307 $\pm$ 64          & \BsKKsuBdKpidef\   &  0.324 $\pm$ 0.019 $\pm$ 0.041   & 24.4 $\pm$ 1.4 $\pm$ 4.6   \\
\hline
\BsKpi         & 230 $\pm$ 34 $\pm$ 16  & \BsKpisuBdKpidef\  &  0.066 $\pm$ 0.010 $\pm$ 0.010   & 5.0 $\pm$ 0.75 $\pm$ 1.0   \\
               &                        &  \ACPsdef\         &  0.39 $\pm$ 0.15 $\pm$ 0.08      &                            \\
               &                        &  \rateratiodef\    &  -3.21 $\pm$ 1.60 $\pm$ 0.39     &                            \\
\Bspipi        & 26 $\pm$ 16 $\pm$ 14   &\BspipisuBdKpidef\  &  0.007 $\pm$ 0.004 $\pm$ 0.005   & 0.53 $\pm$ 0.31 $\pm$ 0.40 \\
	       &                        &      		     &  		                & ($< 1.36$ @~90\%~CL)       \\
\BdKK          & 61 $\pm$ 25  $\pm$ 35  & \BdKKsuBdKpidef\   &  0.020 $\pm$ 0.008 $\pm$ 0.006   & 0.39 $\pm$ 0.16 $\pm$ 0.12 \\
	       &                        &      		     &  	                        &  ($< 0.7$ @~90\%~CL)       \\
\LbpK          & 156 $\pm$ 20 $\pm$ 11  &\LbppisuLbpKdef\    &  0.66 $\pm$ 0.14 $\pm$ 0.08      &                            \\
\Lbppi         & 110 $\pm$ 18 $\pm$ 16  &                    &                                  &                            \\
\br
\end{tabular}
}
\end{table}
\end{center}
The relative branching fractions 
are listed in Table~\ref{tab:summary}, where $f_{d}$ and $f_{s}$ indicate
the production fractions respectively of \Bd\ and \Bs\
from fragmentation of a $b$ quark in $\bar{p}p$ collisions.
An upper limit is also quoted for modes in which no significant signal is
observed. We also list absolute results obtained by normalizing the data to 
the world--average of \BR(\BdKpi)~\cite{HFAG06}. 

The branching fraction of the newly observed mode 
$\BR(\BsKpi)=(5.0 \pm 0.75 \pm 1.0)\times 10^{-6}$ is in agreement with the latest
theoretical expectation, \cita{zupan} which is lower than  the previous predictions \cite{B-N,Yu-Li-Cai}.
We measured for the first time in the \Bs\ meson system
the direct \CP\ asymmetry $\acpbskpi=0.39 \pm 0.15 \pm 0.08$.
This value favors a large \CP\ violation in  \Bs\ meson decays, 
although it is also compatible with zero. 
In Ref.~\cite{Lipkin-BsKpi}  a robust test of the Standard Model or a probe of new physics is suggested by
comparison of the direct \CP\ asymmetries in  \BsKpi\ and \BdKpi\ decays.
Using HFAG input~\cite{HFAG06} we measure  $\frac{\Gamma(\aBdKpi)-\Gamma(\BdKpi)}{\Gamma(\BsKpi)-\Gamma(\aBsKpi)}  
= 0.84 \pm 0.42 \pm 0.15$, 
in agreement with the Standard Model expectation of unity. Assuming that the relationship above
is equal to one, using as input the \BR(\BsKpi) measured here, and the world--average 
for \acpbdkpi\  and  \BR(\BdKpi)~\cite{HFAG06}, 
we estimate the expected value for
$\acpbskpi \approx 0.37$ in agreement with our measurement. 
The branching fraction $\BR(\BsKK)= (24.4 \pm 1.4 \pm 4.6) \times 10^{-6}$ is in agreement with the 
latest theoretical expectation~\cite{matiasBsKK,matiasBsKK2}
and with the previous CDF measurement \cite{paper_bhh}.
An improved systematic uncertainty is expected for the final analysis of the same sample.
The results for the \Bd\ are in agreement with world--average values~\cite{HFAG06}.
$\acpbdkpi =-0.086 \pm 0.023 \pm 0.009$ is already competitive 
with the current \babelle\ measurements.\\
With full Run II samples ($5-6$~\lumifb\ by year 2009) we expect  a measurement of \acp\ in \BdKpi\ 
with a statistical plus systematic uncertainty at 1\% level; 5-sigma observation 
of direct \acp\ in \BsKpi\ (or alternatively the possible indication of non-SM sources of CP violation);
the first measurement of  \acp\ in \Lb\ charmless decays; and 
improved limits, or even observation, of annihilation modes \Bspipi\ and \BdKK.
In addition to the above, time-dependent measurements will be performed for \Bdpipi\ and \BsKK\ decay \cita{CKM06_punzi}. 
See \cite{beauty06_morello,my_thesis} for more details.


%
%

\section{First observation of \BsDsK}
For the search of \BsDsK\ mode we used a unbinned Maximum Likelihood fit which combines the kinematic and  
particle identification information, as the \Bhh\ analysis. The variables used are the invariant mass 
of the \Bs\ candidates in the $D_{s}\pi$ hypothesis and the \dedx\ of the \Bs\ daughter track. 
With a data sample of  integrated luminosity  $\int\Lumi dt\simeq 1.2$~\lumifb\ 
we observe for the first time \BsDsK\ decays,
with a yield of $109\pm19$ events corresponding to a statistical significance of $7.9\sigma$.
We measured its branching fraction
normalized to \BsDspi\ mode:
\begin{equation}
\frac{\BR(\BsDsK)}{\BR(\BsDspi)}=0.107 \pm 0.019\stat\ \pm 0.007\syst.
\end{equation} 
This is the initial step for a possible time-dependent asymmetry measurement
with full Run II statistics. See \cite{bsdsk_webpage,bsdsk_public_note} for more details.

\end{document}